\documentclass[conference]{IEEEtran}
\IEEEoverridecommandlockouts
\usepackage{cite}

\AtBeginDocument{%
  \providecommand\BibTeX{{%
    \normalfont B\kern-0.5em{\scshape i\kern-0.25em b}\kern-0.8em\TeX}}}

\usepackage{amsmath}
\usepackage{algorithm}
\usepackage{algpseudocode}
\usepackage{graphicx}
\usepackage{textcomp}
\usepackage{xcolor}

\begin{document}



\author{\IEEEauthorblockN{Mark Horeni}
\IEEEauthorblockA{\textit{Computer Science and Engineering} \\
\textit{University of Notre Dame}\\
Notre Dame, USA \\}
\and
\IEEEauthorblockN{Siddharth Joshi}
\IEEEauthorblockA{\textit{Computer Science and Engineering} \\
\textit{University of Notre Dame}\\
Notre Dame, USA \\}
}

\title{Improvements in Interlayer Pipelining of CNN Accelerators Using Genetic Algorithms
}

\maketitle

\begin{abstract}
Deploying Convolutional Neural Networks (CNNs) on edge platforms necessitates efficient hardware acceleration. Any unnecessary data movmement incurred in such accelerators can unacceptably degrade performance and efficiency. To address this, we develop a layer fusion technique targeting CNNs, that reduces off-chip data communication using a Genetic Algorithm (GA) applied to graph-based topological sort. Results show a 1.8$\times$ increase in energy efficiency and 1.9$\times$ improvement in energy-delay product (EDP) for MobileNet-v3 on a SIMBA-like mobile architecture. Our approach consistently improves workload performance, averaging 1.4$\times$ improvement to EDP for SIMBA and 1.12$\times$ for Eyeriss. 
\end{abstract}


\section{Introduction}

The widespread deployment of machine learning (ML) on mobile and edge systems has necessitated the development of specialized domain-specific accelerators~\cite{edgetpu, eyeriss, simba}. In particular, accelerators for virtual and extended reality, robotics, and mobile autonomous applications must remain performant while operating under stringent constraints, such as sub-millisecond latency and micro --- milliwatt power consumption. These systems often employ Deep Convolutional Neural Networks (CNNs)~\cite{resnet, unet}, which can be quite energy intensive, requiring cooptimization of both the hardware and algorithms to be feasibly deployed. 

\begin{figure}
    \centering
    \includegraphics[width=.9\columnwidth]{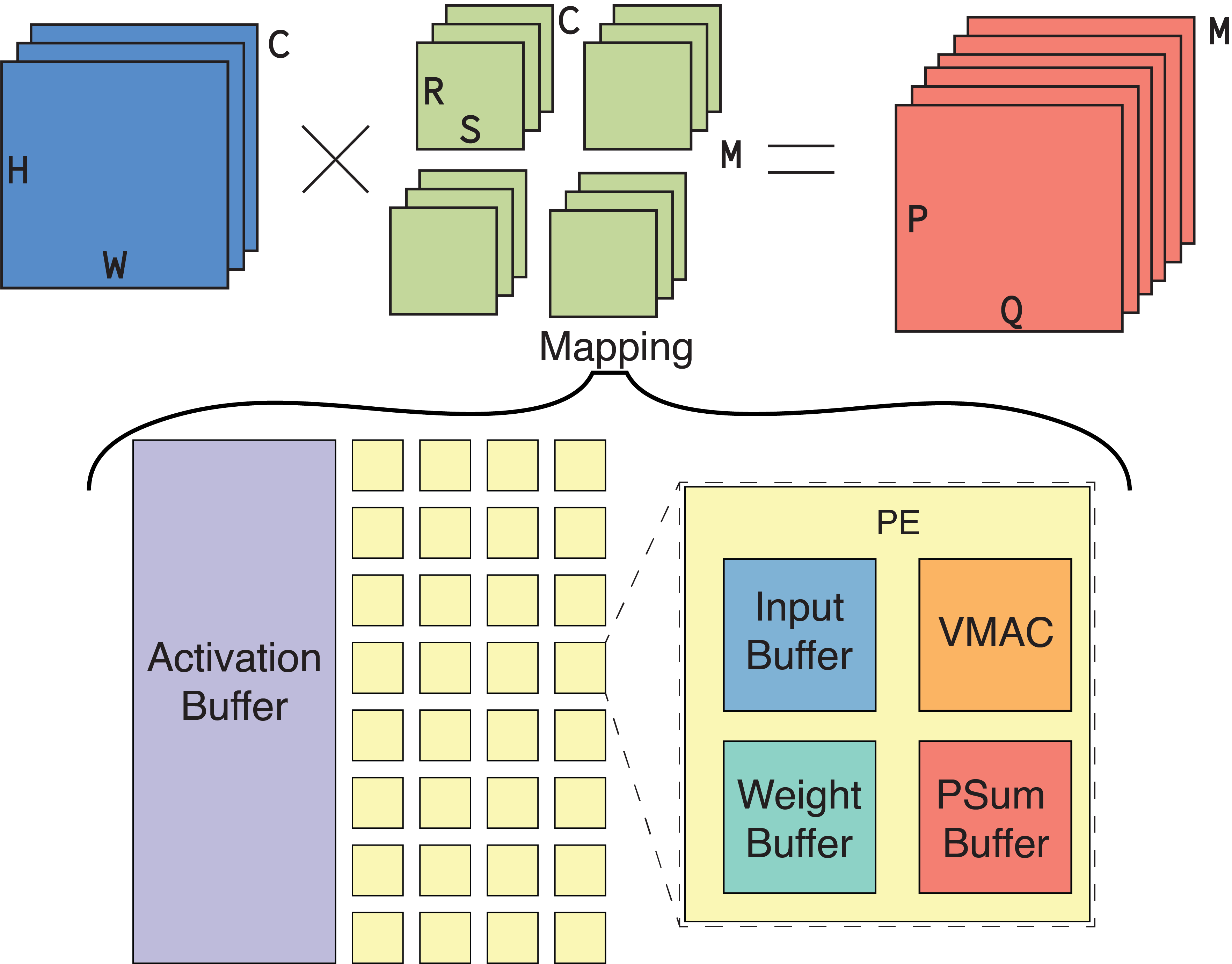}
    \caption{Mapping a convolutional layer in a CNN onto an accelerator, involves allocating memories across multiple levels in the hierarchy to different tensor tiles. Here, blue inputs are multiplied and added to the green weights to produce the red outputs. A template domain-specific accelerator that includes an activation buffer and an array of Processing Elements (PEs), each with its dedicated local scratchpad and a vector multiply-accumulate (VMAC) unit.}
    \label{fig:mapping}
\end{figure}

Codesign between the ML models and accelerators is crucial to achieving energy-efficient operation, given the ever-increasing complexity of ML models~\cite{mobilenetv3, wu2022tinyvit}. Implementing such codesign requires rapid performance evaluation (power, performance, and area --- PPA) and design-space exploration. Tools for such codesign usually comprise three broad subcomponents: an architecture cost model, mapspace generation and exploration, and an architectural template~\cite{timeloop,zigzag}. This approach can capture the complex interaction between accelerator design choice and data movement, enabling rapid pre-RTL design iterations and optimizations~\cite{accelergy}. These tools optimize data movement and communication patterns through different tensor allocation schedules for memories across the hierarchy that might optimize for some target objective. Each schedule results in a mapping, e.g., Fig.~\ref{fig:mapping}, where one set of tensor allocations from a CNN workload is shown for a spatial processor array consisting of an array of Processing Elements (PEs) and a unified activation memory. The input feature map of dimensions C$\times$H$\times$W (where C is the number of channels, H is the height, and W is the width) is convolved with M filters of size C$\times$R$\times$S (where R and S are the height and width of the filter), resulting in an output feature map of size M$\times$P$\times$Q (where P and Q are the height and width of the output feature map). 

\begin{figure}
    \centering
    \includegraphics[width=.9\columnwidth]{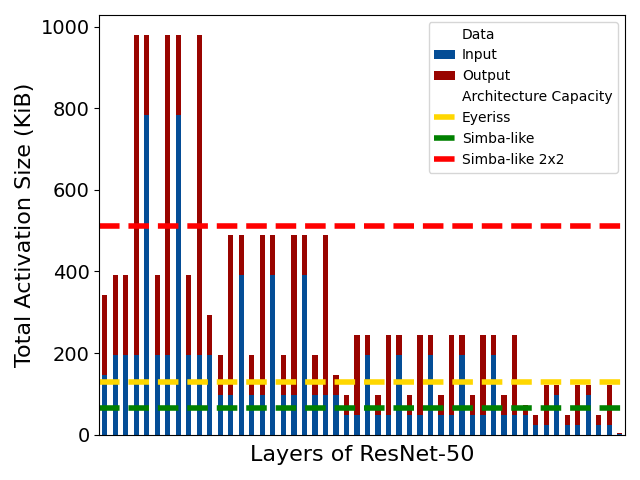}
    \caption{Comparison between input and output activation footprints for different layers in ResNet-50. The total capacity of different architectures is highlighted.}
    \label{fig:needed}
\end{figure}

Mobile and edge accelerators are typically area constrained and cannot store all tensors entirely on-chip, necessitating off-chip communication as shown in Fig.~\ref{fig:needed}. Accelerator dataflow, i.e., data movement across hierarchies and PEs and tensor tiling and allocation to on-chip memoriesis key to performance. However, evaluating this performance entails rapidly navigating a complex space of possible dataflows, with complexity increasing combinatorially with memory hierarchy, number of operand tensors, and parallelism. As a result, most mapping optimization is limited to per-layer exploration of the mapspace. Each layer in a neural network is optimized independent of the prior or subsequent ones. However, multi-layer optimizations can minimize the storage and generation of intermediate tensors, dramatically improving the system’s energy efficiency and latency~\cite{dnnfuser,defines,tvm}.





This work enhances Timeloop~\cite{timeloop} a layer-wise mapping tool, to efficiently optimize data movement multi-layer schedules for CNNs~\cite{fusedlayercnns}. We use a Genetic Algorithm (GA), leveraging topological sort to enforce dependency in the computational graph. This allows us to optimize the use of on-chip resources and minimize tensor offloading to DRAM. We evaluate our tool over multiple architectures for different CNNs, observing performance improvements when evaluated on both energy and the energy-delay product (EDP). Finally, using our tool, we conduct an architectural analysis and demonstrate that a repartitioning repartition tensor allocation to on-chip memories in a multi-layer optimization-aware fashion can further improve architectural performance of EDP by 1.2$\times$.

\section{Background}

\begin{figure}
    \centering
    \includegraphics[width=.9\columnwidth]{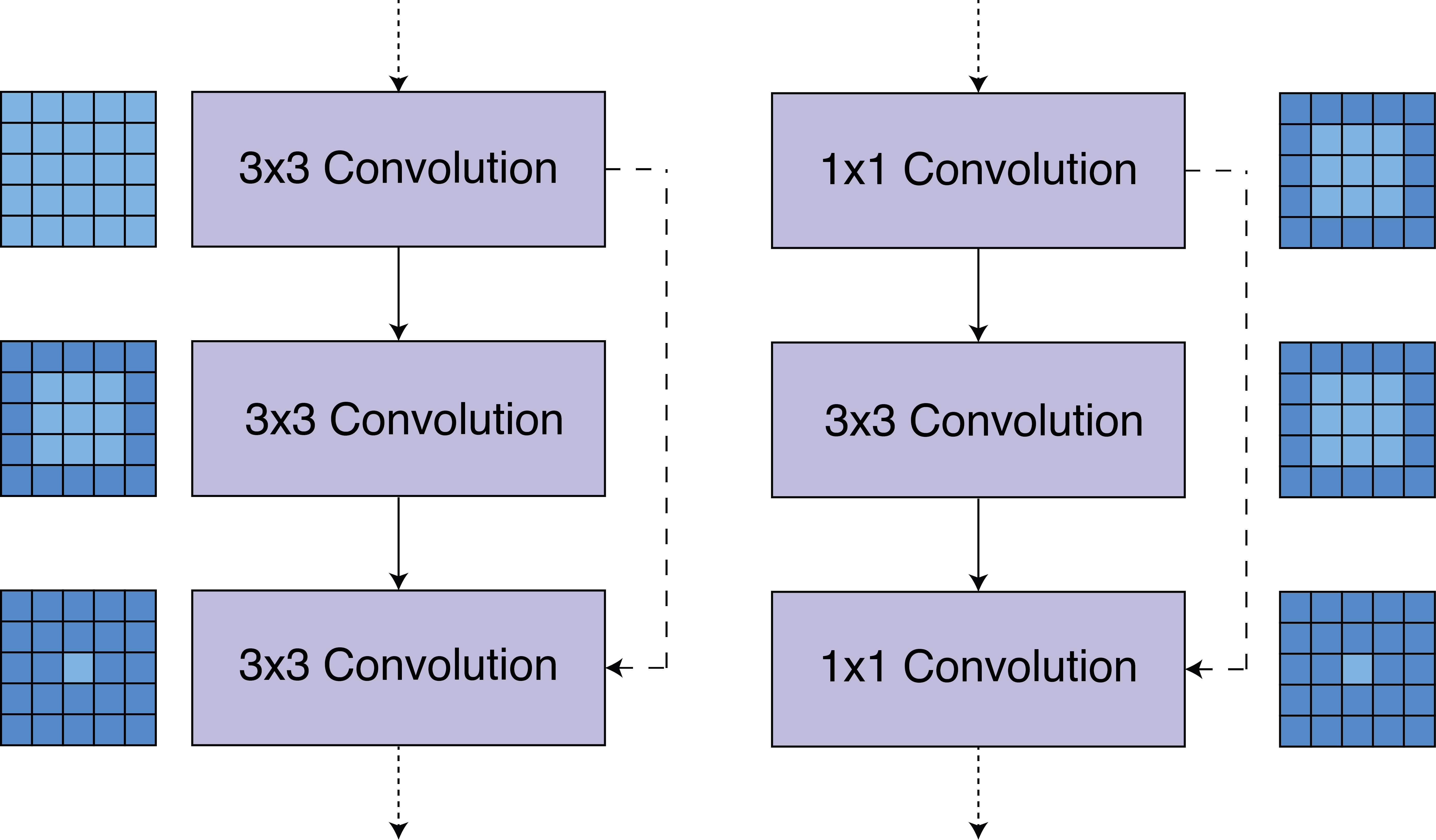}
    \caption{Example of residual blocks within a CNN. These blocks are executed in sequence, with a skip connection where the output of one layer becomes the input of the next layer. We show the difference between the receptive field of a 3$\times$3 convolution (left) grows differently from the receptive field of a pointwise convolution (right).}
    \label{fig:resnetblocks}
\end{figure} 

CNN computational graphs process data sequentially, with each layer weighing the inputs, and applying a nonlinearity to produce outputs that serve as inputs for subsequent layers (e.g., Fig.~\ref{fig:resnetblocks}). Modern CNNs employ more complex topologies, such as skip connections or residual layers that can incur complex long-range dependencies across layers, as depicted in Fig.~\ref{fig:resnetblocks}. Such models still have multiple convolution blocks processing data in sequence while recirculating parts of the outputs to later stages of the network.

\subsection{Accelerators and Mapping}

Common accelerator architectures use arrays of PEs with access to fast, local, dedicated memories. These dedicated memories are typically organized based on tensor types, such as inputs, outputs, and weights. Local communication between PEs, staged memory hierarchies, and broadcasts are common methods to minimize data movement and amortize the access costs of various tensors. Further up the memory hierarchy, memories serve larger numbers of compute units while incurring a higher access cost, with the highest cost being DRAM. Scheduling (or mapping) tensor operations onto hardware involves complex optimizations that allocate memory and compute resources to different tensors while maximizing efficiency and ensuring correctness. 

In this paper, we evaluate two common baseline architectures, a Simba-like \cite{simba} and an Eyeriss-like \cite{eyeriss} architecture modeled and evaluated using the Accelergy+Timeloop suite\cite{accelergy,timeloop}. Our modeling is based on publicly available calibrated models\footnote{https://github.com/Accelergy-Project/baseline-designs/}.

\subsection{Vertical Scheduling for Neural Networks}

\begin{figure}
    \centering
    \includegraphics[width=.95\columnwidth]{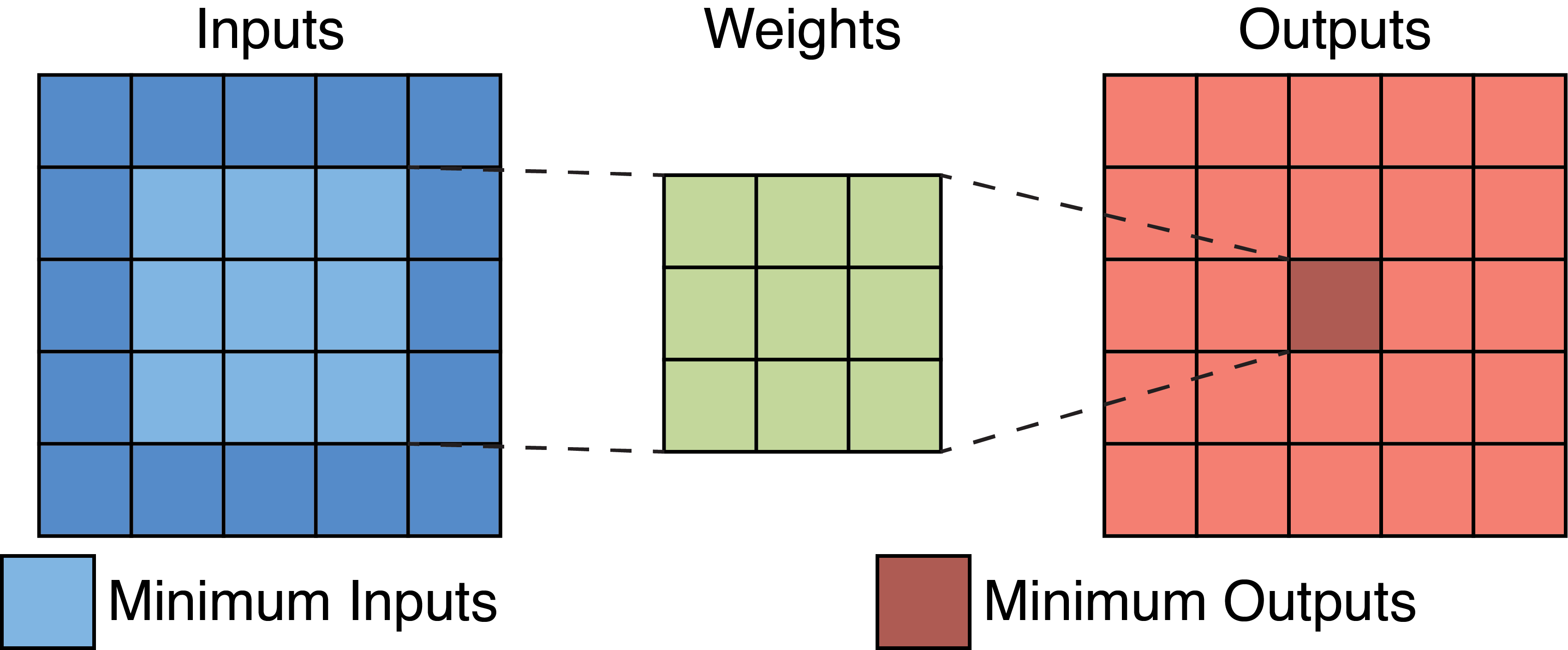}
    \caption{The receptive field of a $3\times3\times1$ CNN filter.}
    \label{fig:pyramid}
\end{figure} 

Fusing operations and scheduling multiple layers are critical to improving accelerator performance~\cite{fusedlayercnns, geneticfusion,dnnfuser,optimus,defines}. Consider an MLP with a single vector input. The input will be multiplied by the weights, followed by applying a nonlinearity before subsequent layers can use it. Reducing the data movement overhead by never storing intermediate values leads to the energy savings associated with vertical scheduling. These gains can further increase convolutional layers at the cost of increased complexity in scheduling. As shown in Fig.~\ref{fig:pyramid} (and Fig.~\ref{fig:resnetblocks}), given a 3$\times$3 window, a single output needs at least 9 inputs to be computed before the application of the nonlinearity and data is available for the next layer(s). This requires the dedicated on-chip memory to hold previous, intermediate, and output feature maps. Additional costs are incurred for caching or recomputing partially calculated values that can be reused for the receptive field calculations. While recomputing is an option, previous works have found that caching is almost always better, so we employ that for future evaluations \cite{fusedlayercnns,dnnfuser,defines}.

\begin{figure}
    \centering
    \includegraphics[width=.95\columnwidth]{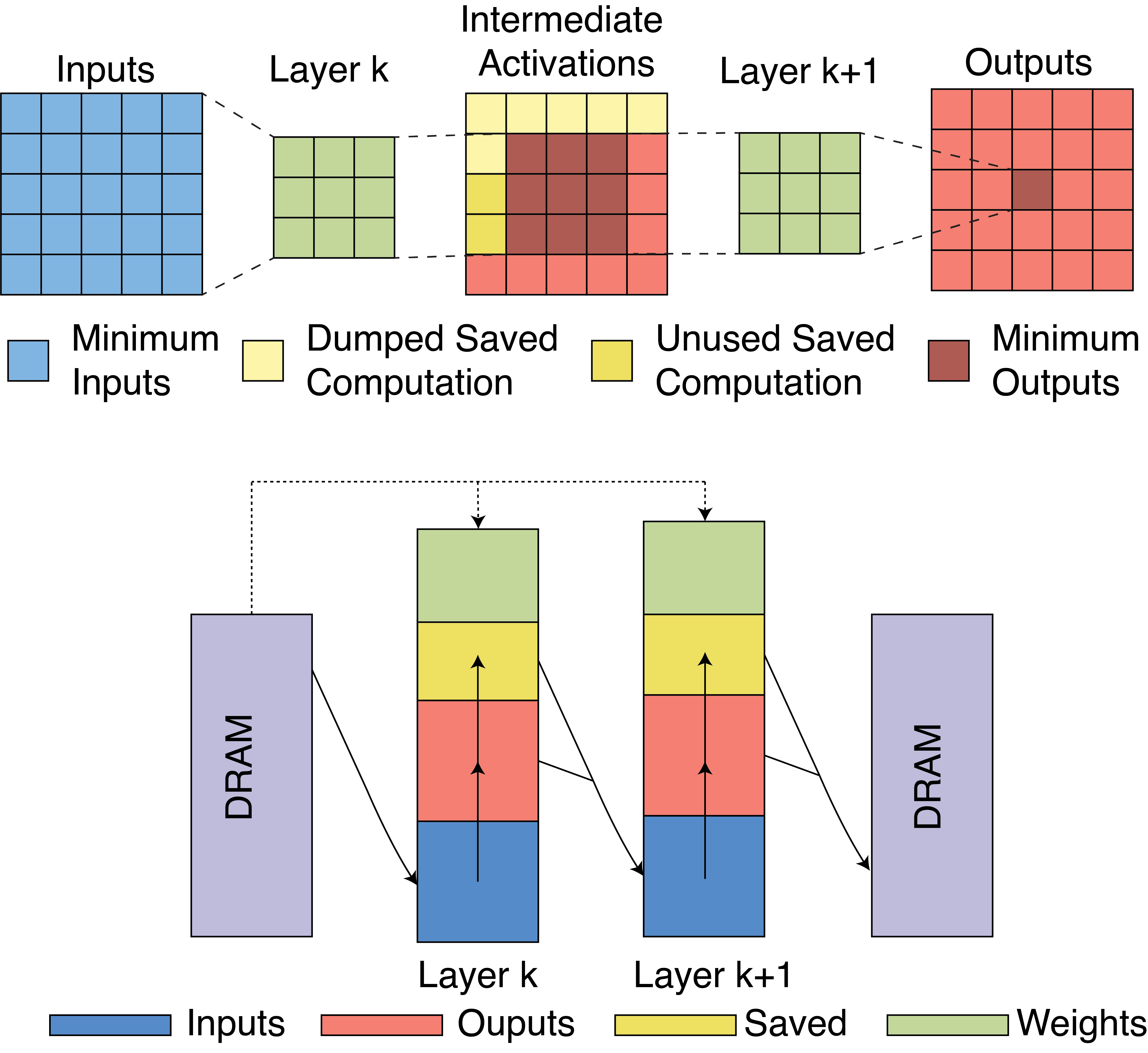}
    \caption{A two-layer network and its receptive field. To understand this, we backtrace from the output layer towards the input. The middle pixel on the output layer depends on inputs from the first layer. All intermediate inputs are required to produce the final output pixel. The intermediate values that must be saved for future operations are highlighted in gold, while those that can be discarded are highlighted in light yellow. One example of data movement for this architecture is shown below. Here, values are read from DRAM and offloaded to DRAM in the case of insufficient on-chip memory. The inputs from one layer generate outputs that will serve as inputs for the subsequent layer. The outputs are then offloaded to DRAM if needed.}
    \label{fig:pyramid_allocation}
\end{figure}

\begin{figure}
    \centering
    \includegraphics[width=.95\columnwidth]{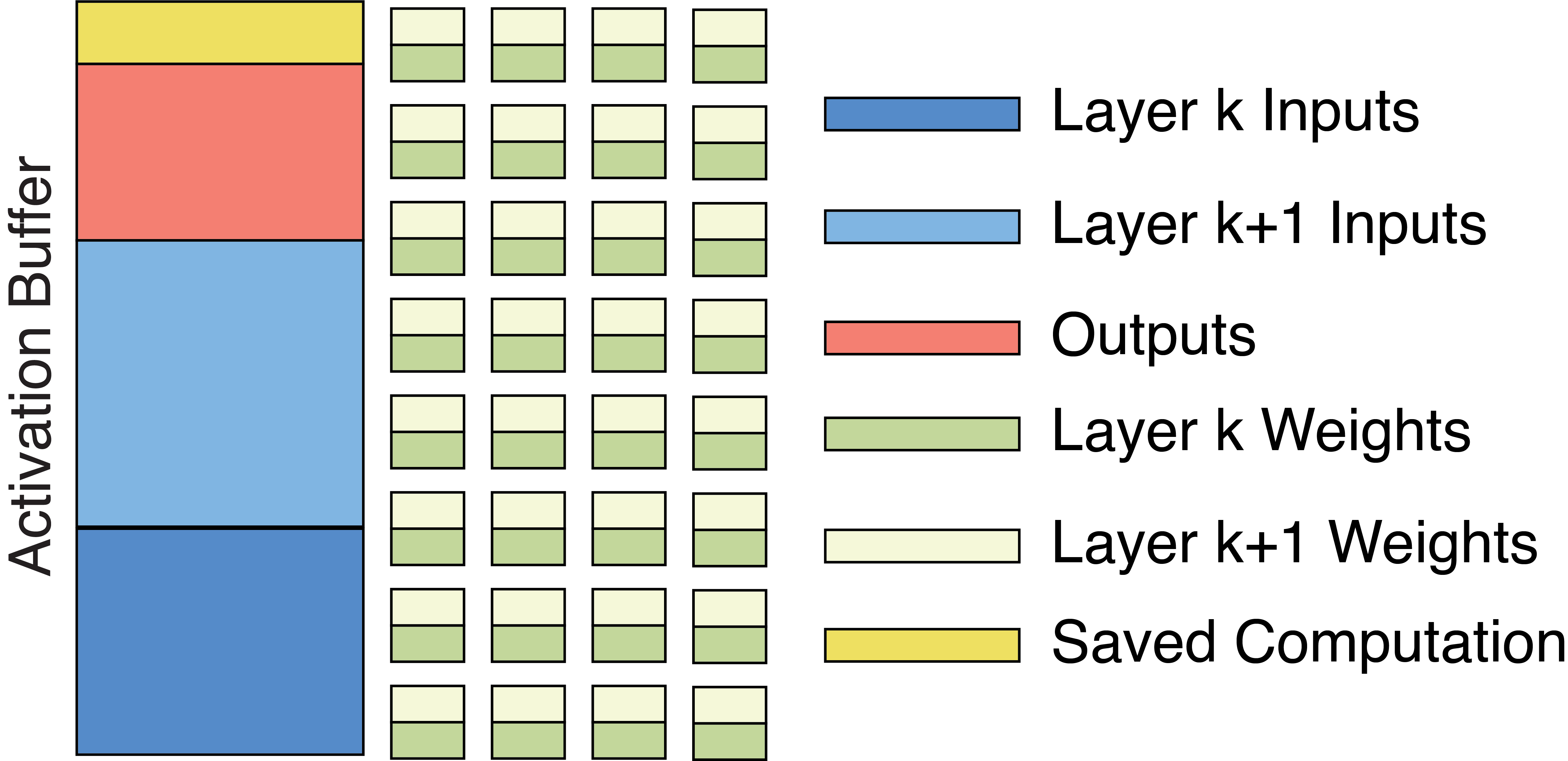}
    \caption{Example allocation of an accelerator. On the left side, the activation buffer is filled with the inputs of layers k and k+1 and the outputs and saved intermediate computations. Meanwhile, inside the PEs, the weights needed for processing are distributed. This allows for the inputs of layer k to be distributed to each PE to form the inputs of layer k+1. After the inputs of layer k+1 are formed, they can be sent back to the PEs to obtain our final outputs.}
    \label{fig:architecture_allocation}
\end{figure}

Figure ~\ref{fig:pyramid_allocation} captures this interaction for a two-layer schedule. In this illustration, these tensors, including partial results, can be stored on-chip without DRAM access. Inputs and weights for a receptive field are accessed from DRAM, with multiple layers (k and k+1) processed before the output of layer k+1 is stored in DRAM.
To determine the central pixel following layer k+1, we require the 9 central pixels from layer k's output, using the full input feature map. The bottom-left pixel corresponding to the nine central pixels is computed to process the entire input feature map. This pixel is used for both the center and to determine subsequent pixels in the final output feature map. Once used completely, the result of previously stored computations can be discarded, enabling the reallocation of storage for upcoming computations. An illustrative mapping for this on an accelerator is shown in Fig.~\ref{fig:architecture_allocation}.

\section{Interlayer Mapping Optimization}
\begin{figure}
    \centering
    \includegraphics[width=.9\columnwidth]{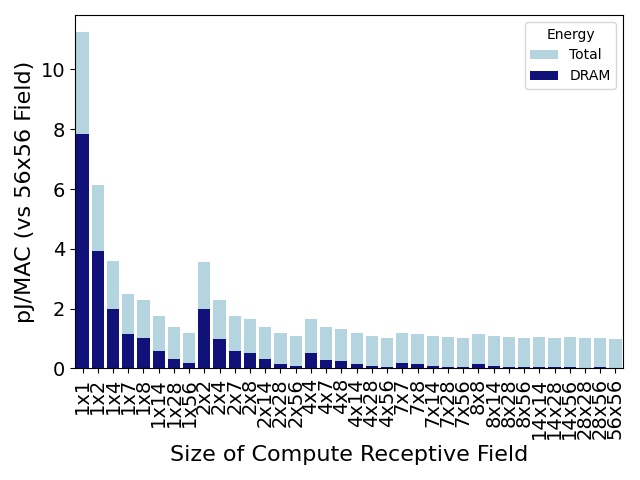}
    \caption{As the size of a feature map increases in the early layers of ResNet-50, more computation affects energy efficiency as measured in pJ/MAC. This measurement shows that we have some relative memory cost for every operation associated with the compute. When more data is loaded into buffers closer to the computation, the memory access is cheaper, allowing for greater energy efficiency.}
    \label{fig:pjmac}
\end{figure}

Previous work, such as \cite{fusedlayercnns}, focused on generating a single output pixel at the bottom of the receptive field. This can be suboptimal since most modern architectures also attempt to maximize operand reuse locally. Only targeting a single output pixel at a time can reduce the energy efficiency, as shown in Fig.~\ref{fig:pjmac}. This figure shows the normalized energy per operation when evaluated using~\cite{timeloop,accelergy} for a Simba-like architecture on an early layer of ResNet-50. The total feature map size is 56$\times$56. We evaluate this for all possible sizes of the receptive field. This provides insight into the reuse pattern and the impact of reloading tensors multiple times. Larger receptive fields can amortize the cost of each memory access across more computation, indicating a balance between receptive field size and layer fusion depth. 

\subsection{Space of Layer Fusion Operations}
First, we will clarify the notation. Consider three layers, such that k+2 depends on k+1, which in turn depends on k. If all three layers can avoid access to off-chip data, we denote them as being \texttt{fused}. This is illustrated in Fig.~\ref{fig:space}. 
Similarly, if layer fusion is impossible and some off-chip access must occur, we denote those layers as being \texttt{split}. 
Given that \texttt{split} and \texttt{fused} are mutually exclusive states for model layers, the scheduling state-space for an accelerator will be exponential in the number of layers for that model. For instance, a neural network like VGG-16 has a potential state space of size $2^{16}$. More complex model topologies, like ResNets with residual connections, have a larger state space. Similarly, for each additional layer, we have the option to try to create a larger set of \texttt{fused} layers by \texttt{combining} them or to \texttt{split} them by \texttt{separating} them. These actions are shown in Fig.~\ref{fig:space}. On-chip memory capacity typically limits how layers can be fused. However, because reuse can be severely impacted by smaller feature maps being computed, it may be more optimal to go off-chip to fuse more future layers with larger receptive fields than a few layers fused once.
Additionally, accelerators might have different objectives, such as the number of off-chip memory accesses, total energy, latency, or EDP, which can be used to decide to combine or split layers.

\subsection{Genetic Algorithm Based Optimization}
\begin{figure}
    \centering
    \includegraphics[width=.8\columnwidth]{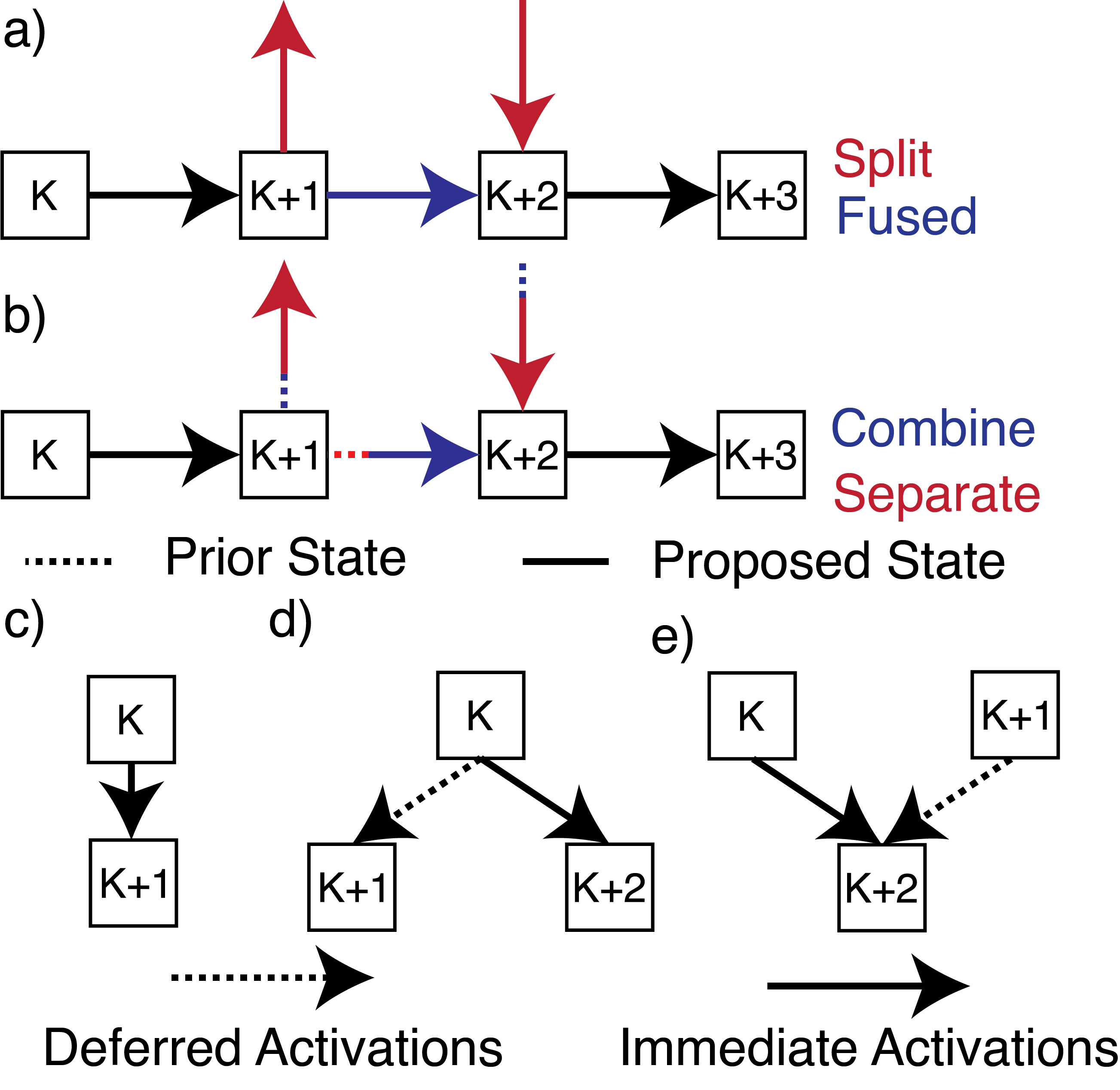}
    \caption{Examples of our scheduling space (a), operations in that space (b), and some elements that comprise that space (c, d, and e). a) Shows an example scheduling state where layer k is directly followed by layer k+1, following k+1 the activation tensors for layers k+1 and k+2 can either be \texttt{split} (shown in red) by offloading activations to DRAM or can be \texttt{fused} (shown in red) by avoiding any off-chip communication for activations. b) Shows an example action space where a previously \texttt{split} layer (dotted blue line) that had some activation tensors offloaded (entirely or partially) can take the action \texttt{combine} to become a fused layer (solid blue arrow) with activation tensors entirely contained on-chip, resulting in fewer memory accesses. The solid and dotted red arrows depict the inverse action (\texttt{separate}). c) The simplest connection type, where a layer has a single input and output.d) A layer can output to layers further in the network where activations would need to be stored if fused. e) A layer can depend on multiple other layers, including layers not evaluated immediately before the current layer.}
    \label{fig:space}
\end{figure}\vspace{-.05cm}

We develop a parallel GA-based approach to optimize the layer-fusion space by determining when layers should be fused or split. A more formal description of our algorithm is provided in Alg.~\ref{alg:full}. We begin by initializing our population with a layer-by-layer schedule. The GA is then allowed to\texttt{fuse} layers by \texttt{combining} them or \texttt{split} them by \texttt{separating} them as shown in Fig.~\ref{fig:space}b. After the population is modified, we calculate a fitness score. We use an incremental improvement fitness score ($F$) as 
    $F = \text{Eval}_\text{Layerwise} /\text{Eval}_\text{New}$.

The metric used for evaluation for the rest of the paper is EDP as it provided the most useful information of the balance between performance and energy improvement.
We also choose the Top-$N$ and a few random mappings to go onto the next generation to ensure we do not quickly converge to a poor local minimum. 

We choose receptive field sizes that maximally use the activation buffer. Here, small but deep receptive fields must be balanced with larger, shallower receptive fields. Trading-off reduced intermediate activation storage against per-layer operand reuse. Figure~\ref{fig:pjmac}, evaluated on a layer of ResNet-50 indicates that larger receptive fields can increase energy efficiency. Given a mapping, we evaluate the largest receptive field for that model layer. Any mapping where intermediate storage exceeds capacity is discarded as invalid.


\subsection{Topological Sort}
We use a topological sort within our GA to ensure that long-range dependencies across CNN layers are accounted for. To do so, we represent our network as a computation graph, with the fused layers being subgraphs. We then determine the scheduling order within the subgraph while maintaining their connection to the main graph to evaluate the cost of the complete model correctly. 

Layers might receive inputs from the immediately preceding layer and an earlier layer in the network. An accurate schedule requires that both intermediate outputs for two different layers be stored until they can be combined. Alternatively, multiple layers can be fused while activations previously offloaded are brought on-chip. This structure is commonly seen in residual blocks, which create long-range dependencies that must be tracked across blocks. The converse scenario (inverted bottleneck) can also occur, as shown in Fig.~\ref{fig:space}d. Here, a layer has outputs to an immediate successor layer and outputs (possibly multiple) to a layer further down the network. This occurs frequently in U-Net~\cite{unet}. A topological sort implemented on all subgraphs created after fusion enforces dependencies in the computational graph. Because not all topological sorts will be unique, we select a random primary graph and its corresponding elements of the subgraph to process. Figure~\ref{fig:space} c---e shows exemplar subgraphs that our approach can process.



\begin{algorithm}
\caption{Multilayer Scheduling Optimization}\label{alg:full}
\begin{algorithmic}[1]
\State \textbf{Initialize:} Entire population $P$ with every layer $K$ and set of splits $S$  starting condition of Layerwise evaluation ($\text{E}_\text{Layerwise})$.
\For{$G$ Generations}
    \For{$C$ Mutations to the Population}
        \State Choose two adjacent layers ($K$ and $K+1$) with $S$ from $P$
        \Statex \hskip\algorithmicindent\hskip\algorithmicindent\hskip\algorithmicindent and update $S$ by choosing \texttt{separate} or \texttt{combine}.
        \State Create a set of subgraphs $G$ based on $S$ subgraphs
        \Statex \hskip\algorithmicindent\hskip\algorithmicindent\hskip\algorithmicindent for each set of connected layers, making sure
        \Statex \hskip\algorithmicindent\hskip\algorithmicindent\hskip\algorithmicindent each subgraph is weakly connected.
        \State Topologically sort all of $G$ to reveal dependencies.
        \State Calculate maximum Receptive Field Size for each
        \Statex \hskip\algorithmicindent\hskip\algorithmicindent\hskip\algorithmicindent $G$, and check the capacity of the weight buffer.
        \State Evaluate and sum each $K$ given $S$ to obtain $\text{E}_{\text{new}}$
        \State Fitness = $\text{E}_{\text{Layerwise}}/\text{E}_{\text{New}}$
        \State Add $\text{E}_{\text{new}}$ to $P$
    \EndFor
    \State Set $P$ as Top-N fitness scores and some random scores.
\EndFor
\end{algorithmic}
\end{algorithm}


\subsection{Baseline Evaluation}
\begin{figure}
    \centering
    \includegraphics[width=.9\columnwidth]{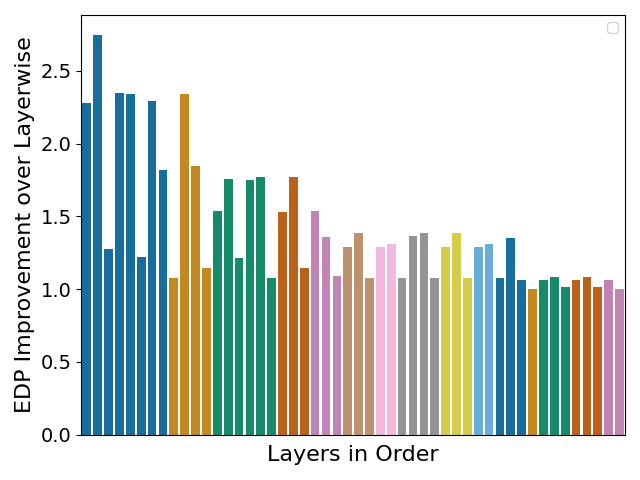}
    \caption{Visualization of the fused groups of ResNet-50 on our Simba-2x2 architecture. Adjacent bars with the same color are fused and have activations remain on-chip.}
    \label{fig:layers}
\end{figure}

\section{Evaluation}

We demonstrate the versatility of our approach by evaluating scheduling performance on multiple CNN models and accelerator architectures (see Table~\ref{tbl:arch}). We evaluate fused and unfused layers using Timeloop+Accelergy~\cite{accelergy,timeloop}. We also modify the Eyeriss architecture to include an intermediate weight buffer of 512 KiB, equal to that of a single SIMBA chiplet, to store multiple layers simultaneously. This better enables a fair comparison between dataflows since Eyeriss's row stationary dataflow would require more off-chip weight with no benefit from keeping weights on-chip. We also assume all accelerators operate at a nominal clock frequency of 200 MHz (low power mode) connected to an LPDDR4 DRAM with a transfer bandwidth of 128 GB/s. Our GA is configured to use a population size (P) of 100, with the top 10 (N) selected per generation, evaluated over 500 generations (G).

\begin{table}[h]
\centering
\caption{Configuration of Evaluated Architectures}
\begin{tabular}{|l||l|l|l|}
\hline
\textbf{Arch}                       & \textbf{Eyeriss} & \textbf{SIMBA} & \textbf{SIMBA $2\times2$} \\ \hline
\textbf{PEs X}                      & 14               & 4              & 8                  \\ \hline
\textbf{PEs Y}                      & 12               & 4              & 8                  \\ \hline
\textbf{MACs per PE}                & 1                & 64             & 64                 \\ \hline
\textbf{Activation Buffer (KiB)} & 128           & 64          & 256 (Total)     \\ \hline
\textbf{Weight Buffer (KiB)} & 512           & 512         & 2048 (Total)    \\ \hline
\end{tabular}
\label{tbl:arch}
\vspace{-5pt}
\end{table}

We evaluate a 2$\times$2 SIMBA-like architecture as a baseline to explain our methodology. Figure ~\ref{fig:layers}, depicts the automated schedule developed by our tool for a ResNet-50 on a single SIMBA chiplet, resulting in a significant improvement in EDP in earlier layers (up to 2.7$\times$) and overall (1.2$\times$). The best mapping found by our algorithm reduced activations written to DRAM significantly, writing to DRAM 15 times instead of 50 for all layers of ResNet-50. Since the early layers of ResNet-50 have large activiations but few weights,  the first subgraph can easily fit on-chip. However, some weights may be too large for deeper parts of the network, even without fusion. In such cases, they must always be loaded from DRAM, and fusion of activations only helps prevent some unnecessary off-chip memory accesses. While this helps in energy efficiency, it does not impact the number of cycles due to sufficient DRAM bandwidth and Timeloop's schedules creating computation and communication overlap. Some configurations see a performance decrease due to Timeloop's factorization-based mapping, which prevents full array utilization.

\begin{figure*}[ht!]
    \centering
    \includegraphics[width=\textwidth]{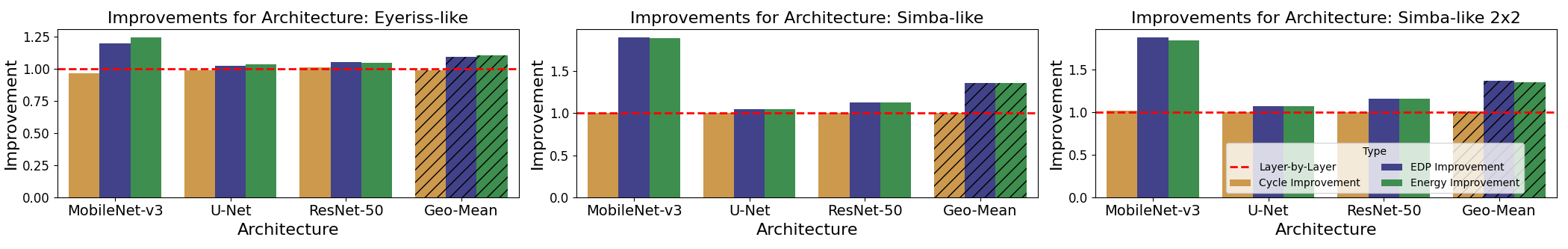}
    \caption{Each graph depicts our separate architecture configurations along with three networks (MobileNet-v3, U-Net, and ReseNet-50) and the geometric mean between them. We see that overall, we have our greatest improvement with MobileNet-v3 and modest gains with U-Net and ResNet-50.}
    \label{fig:morenets}
\end{figure*}
\begin{figure}
    \centering
    \includegraphics[width=.9\columnwidth]{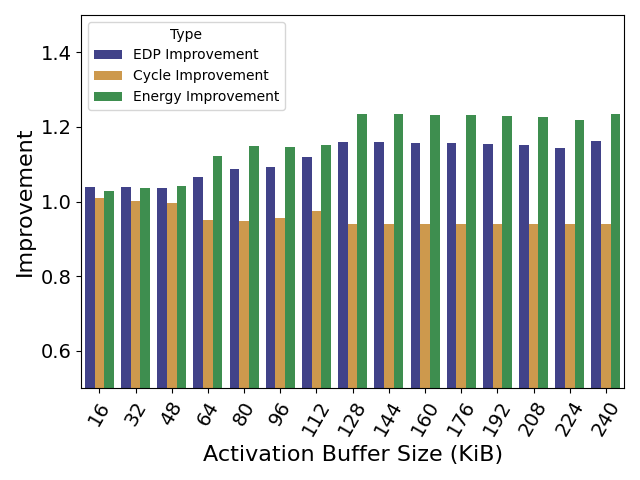}
    \caption{From left to right, we increase or decrease the activation buffer memory at the expense of the weight memory in our Eyeriss-like configuration on ResNet-50. Some later configurations have increased energy efficiency but lack performance improvement due to Eyeriss's 14$\times$12 array and the inability to utilize the array completely.}
    \label{fig:ratio}
\end{figure}
\subsection{Architecture Modifications}

Figure~\ref{fig:ratio}, examines the balance in memory allocation between weights and activations for the Eyeriss-like architecture at iso-memory capacity. We trade off 16 KiB of memory between the weight buffer and activation buffer in our Eyeriss evaluation and provide a breakdown in Energy, Cycles, and enhancements EDP. To accommodate more activations on-chip, we increase the activation capacity. Because Eyeriss employs a row-stationary dataflow, this dataflow maximizes on-chip data movement. Due to reducing memory allocation to weight tensors, Eyeriss can better minimize movement of the activation tensors. Figure ~\ref{fig:morenets} summarizes the performance for different CNN workloads across our evaluated architectures. Our findings indicate that the shallower networks with large activation sizes and fewer weights benefit most from such scheduling. Performance benefits can be attributed to storing weights from multiple layers on-chip and calculating larger receptive fields while minimizing offloads. However, larger layers prevent effective reuse of weights due to the need to reload offloaded weight tensors constantly. This is exacerbated for the depthwise separable layers in MobileNet-v3, which have a high ratio between the size of activation and weight tensors.


\section{Related Work}
Mapping optimization for ML models includes operator fusion and layer fusion\cite{defines, optimus, geneticfusion, looptree}. Optimus \cite{optimus} focuses specifically on operator fusion (i.e., fusing Conv + Act + BatchNorm) on general directed computational graphs to reduce memory accesses given dependencies; they enforce operation order using a topological sort. Since they only look at operator fusion, they only evaluate a limited number of nodes in their computational graph. \textit{DEFINES} \cite{defines} minitmizes data-copy through a depth-centric mapping. This greedily maximizes layerwise depth, while neglecting other off-chip data movement and intra-layer operand reuse for weight and activation tensors. We aim to maximize the receptive field at each layer split/fusion to maximize reuse and incorporate any additional data movement required for weights and activations. GAs have been studied for workload allocation across different cores~\cite{geneticfusion}. However, their work uses the GA to split activation tensor along a single dimension only, rather than using the receptive fields. 

On data center accelerators, layer fusion typically evaluates batch splitting~\cite{tangram, dnnfuser}. This is typically not feasible for edge inference applications where only one or two batches (stereo vision) might be available. DNNFuser~\cite{dnnfuser} studies mini-batch splitting to limit the mapping options and optimize the sequential decision-making using decision transformers. Tangram~\cite{tangram} enforces a weight stationary dataflow, where multiple CNN layers are allocated on-chip, enabling activations to be streamed across the different PE regions. This avoids any weight-tensor related data-movement entirely but might not be feasible for edge systems. Recent work~\cite{fusedlayercnns} investigated an architecture optimized to maximize the receptive fields computed on-chip, however, their approach is specific to their architecture and results in sub-optimal performance on SIMBA-like architectures for deeper layers in a CNN.

\section{Conclusion}
We introduced a new tool that optimizes how CNN models are deployed for inference on user-defined edge ML accelerators. Our tool consistently improves model performance across multiple accelerators, indicating an average of 1.4$\times$ energy improvement on the Simba and Simba 2$\times$2 model and a 1.15$\times$ energy improvement on the Eyeriss model. Further studies indicate that modifying the memory allocation in Eyeriss' design delivers a 1.25$\times$ energy improvement over the baseline architecture.
\bibliographystyle{IEEEtran}
\bibliography{refs}

\end{document}